\documentclass[article,preprint,groupedaddress]{revtex4}
\usepackage{epsfig}
\usepackage{graphicx}
\usepackage{amssymb}

\begin{document}

\title{Compact stars: To cross or go around? That is the question}
\author{Shahar Hod}
\affiliation{The Ruppin Academic Center, Emeq Hefer 40250, Israel}
\affiliation{ } \affiliation{The Hadassah Institute, Jerusalem
91010, Israel}
\date{\today}

\begin{abstract}

\ \ \ The travel times of light signals between two antipodal points on the surface of a compact star are calculated 
for two different trajectories: a straight line that passes through 
the center of the star and a semi-circular trajectory that connects the antipodal points along the surface of the star. 
Interestingly, it is explicitly proved that, for highly dense stars, the longer 
trajectory (the one that goes along 
the surface of the star) may be characterized by the {\it shorter} travel time as measured by asymptotic observers. 
In particular, for constant density stars we determine {\it analytically} the critical value of the dimensionless 
density-area parameter $\Lambda\equiv 4\pi R^2\rho$ that 
marks the boundary between situations in which a direct crossing of the star through its center 
has the shorter travel time and situations in which the semi-circular trajectory along the surface of the star is characterized by 
the shorter travel time as measured by asymptotic observers [here $\{R,\rho\}$ are respectively 
the radius of the star and its density].
\end{abstract}
\bigskip
\maketitle

\section{Introduction}

Many of us have encountered situations in which we have to drive from one side of a city to its opposite side. 
Then the dilemma arises: Should I choose the short route that passes through the center of the city, or should I 
choose the longer road that goes along the perimeter of the city?

The answer to this `existential' question obviously depends on the density of traffic (the traffic congestion) that 
characterizes the city in question. 
In particular, one expects that there is a critical traffic load (critical car density) above which it would 
be better for us to choose the longer but less busy road, the one that connects one side of the city 
to its opposite side along the perimeter of the city.

Within the framework of general relativity, time as measured by asymptotic observers in curved spacetimes
is influenced by the spatially-dependent energy density that characterizes the spacetime in question. 
This is one of the fundamental and most known predictions of Einstein's theory of general relativity. 

Suppose a physicist (Alice) has a remote controlled spaceship that can move arbitrarily close to the speed of light and 
she wants to send it from a point A on the surface of a compact star 
to its antipodal point B which is located on the opposite side of the star. 
And suppose Alice can choose between two options: 
(1) To send her remote controlled spaceship from the point A to the point B along a straight line that crosses the star 
directly through its center \cite{Notetun}, or 
(2) To send the spaceship along a semi-circular trajectory that goes along the surface of the star 
from point A to its antipodal point B. 

The following important question naturally arises: Which of these two journeys between the two opposite sides 
of the compact star will take less time as measured by Alice who is located 
far away from the compact star?

The answer to this physically interesting question is quite obvious in the flat-space limit of 
highly dilute stars, in which case the travel time is mainly determined by the length of the trajectory. 
In particular, the straight-line trajectory from a point A to its antipodal point B 
that crosses the star through its center is characterized by the shorter traveling time. 

On the other hand, curved spacetime effects are expected to become important as the dimensionless density 
parameter \cite{Noteunits}
\begin{equation}\label{Eq1}
\Lambda\equiv 4\pi R^2\rho_{\text{max}}\ 
\end{equation}
that characterizes a star of radius $R$ and maximum density $\rho_{\text{max}}$ 
deviates significantly from zero. 
In particular, for highly dense stars, the longer road which takes the remote controlled spaceship 
from the point A to its antipodal point B along a semi-circular trajectory on the surface of the star 
may have the shorter traveling time as measured by the remote operator (Alice). 

The dimensionless density-area parameter $\Lambda$, which characterizes a star of radius $R$ 
and maximum energy density $\rho_{\text{max}}$, is a physically important quantity whose value can 
in principle be bounded from below by far away observers who measure 
the gravitational redshift factor of spectral lines that were emitted from the observed star \cite{Hodzz}. 
In the present compact paper we shall demonstrate that, due to this gravitational 
redshift (time dilation) effect in highly curved spacetimes, the identification of the trajectory that connects 
two antipodal points on the surface of a compact star and has the shorter travel time as measured by asymptotic observers (the remote operator) 
is a highly non-trivial task. 

In particular, for some models of compact stars one may expect to find a 
critical value $\Lambda=\Lambda^*$ for the dimensionless density parameter that 
marks the boundary between situations in which a direct crossing through the center of the star 
has the shorter travel time as measured by asymptotic observers 
and situations in which the semi-circular trajectory along the surface of the star has 
the shorter travel time.

The main goal of the present compact paper is to demonstrate, using an analytically solvable model, 
the existence of a critical value for the dimensionless density-area parameter $\Lambda$ of a compact 
star beyond which the {\it longer} orbit (the semi-circular trajectory along the surface of the star) 
has the {\it shorter} travel time as measured by the remote operator.

\section{Compact stars: To cross or to go around?}

We consider a spatially regular compact star of radius $R$ whose asymptotically flat curved spacetime 
is described, using the Schwarzschild spacetime coordinates, 
by the spherically symmetric line element \cite{Chan,ShTe,May,Hodt1}
\begin{equation}\label{Eq2}
ds^2=-Adt^2 +Bdr^2+r^2(d\theta^2 +\sin^2\theta d\phi^2)\  ,
\end{equation}
where $A=A(r)$ and $B=B(r)$. 

An asymptotically flat spacetime is characterized by the radial functional behaviors \cite{May,Hodt1}
\begin{equation}\label{Eq3}
A(r\to\infty) \to 1\ \ \ \ {\text{and}}\ \ \ \ \ B(r\to\infty)\to 1\
\end{equation}
of the metric functions. 
In addition, a regular spacetime is characterized by the functional relations \cite{May,Hodt1}
\begin{equation}\label{Eq4}
A(r\to0)>0\ \ \ \ {\text{and}}\ \ \ \ \ B(r\to0)\to1\ 
\end{equation}
at the center of the star. We shall assume that the energy density and pressure are zero outside 
the surface $r=R$ of the compact star, in which case the spacetime outside the star is characterized 
by the Schwarzschild line element (\ref{Eq2}) with
\begin{equation}\label{Eq5}
A(r)=[B(r)]^{-1}=1-{{2M}\over{r}}\ \ \ \ \ \text{for}\ \ \ \ \ r\geq R\  ,
\end{equation}
where $M$ is the total (asymptotically measured) mass of the star. 

Our goal is to {\it minimize} the crossing time $T$ of the compact star by the remote controlled spaceship as 
measured by the remote operator (Alice). 
We shall therefore assume that, by using non-gravitational forces, 
the spaceship can move arbitrarily close to the speed of light along a non-geodesic trajectory. 

In this case the crossing time $T_{\text{c}}$ of the star along a straight-line trajectory 
that connects a point A on the surface of the star 
to its antipodal point B and passes through the {\it center} of the star as measured by the asymptotic observers 
can be obtained from the curved line element (\ref{Eq2}) with the properties
\begin{equation}\label{Eq6}
ds=d\theta=d\phi=0\  .
\end{equation}
In particular, substituting Eq. (\ref{Eq6}) into (\ref{Eq2}) one obtains the integral relation  
\begin{equation}\label{Eq7}
T_{\text{c}}=2\int^{R}_{0}{\sqrt{{B(r)}\over{A(r)}}}dr\
\end{equation}
for the travel time through the center of the star. 

On the other hand, the travel time $T_{\text{s}}$ between the two antipodal points along a semi-circular trajectory on 
the {\it surface} of the star as measured by the asymptotic observers 
can be obtained from the line element (\ref{Eq2}) of the curved spacetime 
with the properties
\begin{equation}\label{Eq8}
ds=dr=d\theta=0\ \ \ \ \  \text{and}\ \ \ \ \ \Delta\phi=2\pi\  .
\end{equation}
In particular, substituting Eqs. (\ref{Eq5}) and (\ref{Eq8}) into Eq. (\ref{Eq2}) and performing the azimuthal integration, 
one obtains the functional expression
\begin{equation}\label{Eq9}
T_{\text{s}}={{\pi R}\over{\sqrt{1-{{2M}\over{R}}}}}\
\end{equation}
for the traveling time along a semi-circular trajectory on the surface of the star. 
The travel time (\ref{Eq9}) can be expressed in the dimensionless form
\begin{equation}\label{Eq10}
{{T_{\text{s}}}\over{M}}={{\pi}\over{{\cal C}\sqrt{1-2{\cal C}}}}\  ,
\end{equation}
where ${\cal C}\equiv M/R$ is the characteristic compactness parameter of the star. 

In order to facilitate a fully {\it analytical} treatment of the physical system, 
we shall illustrate our ideas using the analytically solvable model of constant density stars, 
which are characterized by the functional relations \cite{ShTe}
\begin{equation}\label{Eq11}
A(r\leq R)={{1}\over{4R^3}}\Big(3R\sqrt{R-2M}-\sqrt{R^3-2Mr^2}\Big)^2\
\end{equation}
and
\begin{equation}\label{Eq12}
B(r\leq R)=\Big(1-{{2Mr^2}\over{R^3}}\Big)^{-1}\
\end{equation}
for the dimensionless metric functions inside the compact star. 

The metric functions (\ref{Eq11}) and (\ref{Eq12}) can be expressed in a mathematically 
compact form in terms of the dimensionless density parameter $\Lambda$ of the star [see Eq. (\ref{Eq1})]:
\begin{equation}\label{Eq13}
A(x\leq1)={{1}\over{4}}\Bigg(3\sqrt{1-{{2\Lambda}\over{3}}}-\sqrt{1-{{2\Lambda}\over{3}}x^2}\Bigg)^2\
\end{equation}
and
\begin{equation}\label{Eq14}
B(x\leq1)=\Big(1-{{2\Lambda}\over{3}}x^2\Big)^{-1}\  ,
\end{equation}
where we have used here the dimensionless radial coordinate 
\begin{equation}\label{Eq15}
x\equiv {{r}\over{R}}\in[0,1]\  .
\end{equation}

Substituting Eqs. (\ref{Eq13}), (\ref{Eq14}), and (\ref{Eq15}) into Eq. (\ref{Eq7}), 
one finds the integral relation
\begin{equation}\label{Eq16}
T_{\text{c}}=\int^{1}_{0}{{4R}\over{\sqrt{1-{{2\Lambda}\over{3}}x^2}
\Big(3\sqrt{1-{{2\Lambda}\over{3}}}-\sqrt{1-{{2\Lambda}\over{3}}x^2}\Big)}}dx\  .
\end{equation}
Interestingly, and most importantly for our analysis, one finds that the integral (\ref{Eq16}) can be 
evaluated analytically to yield the dimensionless expression
\begin{equation}\label{Eq17}
{{T_{\text{c}}}\over{M}}={{18\Big[\arctan\Big(\sqrt{{{\Lambda}\over{12-9\Lambda}}}\Big)+
\arctan\Big(3\sqrt{{{\Lambda}\over{12-9\Lambda}}}\Big)\Big]}\over{\Lambda^{3/2}\sqrt{12-9\Lambda}}}\
\end{equation}
for the travel time through the center of the star. 

Taking cognizance of the analytically derived expressions (\ref{Eq10}) \cite{Notelmlm} and (\ref{Eq17}) for the 
travel times between the antipodal points of the compact star, one finds that the dimensionless 
ratio $T_{\text{c}}(\Lambda)/T_{\text{s}}(\Lambda)$ is a monotonically increasing function 
of the dimensionless density parameter $\Lambda$ of the star with the asymptotic properties
\begin{equation}\label{Eq18}
{{T_{\text{c}}}\over{T_{\text{s}}}}\to {{2}\over{\pi}}<1\ \ \ \ \ \text{for}\ \ \ \ \ \Lambda\to0\
\end{equation}
and \cite{Notelim,Buch}
\begin{equation}\label{Eq19}
{{T_{\text{c}}}\over{T_{\text{s}}}}\to \infty\ \ \ \ \ \text{for}\ \ \ \ \ \Lambda\to{4\over3}\  .
\end{equation}

The asymptotic functional behaviors (\ref{Eq18}) and (\ref{Eq19}), which characterize the $\Lambda$-dependent dimensionless 
ratio $T_{\text{c}}(\Lambda)/T_{\text{s}}(\Lambda)$, reveal the fact that there must 
exist a critical value $\Lambda=\Lambda^*$ of the dimensionless density-area parameter above which 
the {\it longer} trajectory that connects the antipodal points of the star 
(the semi-circular trajectory that goes along the surface of the star) 
is characterized by the {\it shorter} travel time as measured by the remote operator (Alice). 

The critical density parameter of the constant density stars is defined by the 
functional relation 
\begin{equation}\label{Eq20}
T_{\text{c}}=T_{\text{s}}\ \ \ \ \ \text{for}\ \ \ \ \ \  \Lambda=\Lambda^*\  ,
\end{equation}
or equivalently [see Eqs. (\ref{Eq10}) and (\ref{Eq17})]
\begin{equation}\label{Eq21}
\sqrt{{{9-6\Lambda}\over{\Lambda(12-9\Lambda)}}}
{\Bigg[\arctan\Big(\sqrt{{{\Lambda}\over{12-9\Lambda}}}\Big)+
\arctan\Big(3\sqrt{{{\Lambda}\over{12-9\Lambda}}}\Big)\Bigg]}=
{{\pi}\over{2}}\ \ \ \ \ \text{for}\ \ \ \ \ \  \Lambda=\Lambda^*\  .
\end{equation}
This is a highly non-trivial equation for the critical value $\Lambda^*$ of the dimensionless 
density parameter that characterizes the family of constant density stars. 

At first glance it seems that numerical methods must be used in order to solve Eq. (\ref{Eq21}). 
Interestingly, however, despite the non-linear character of Eq. (\ref{Eq21}), 
the critical density parameter $\Lambda^*$ can be determined {\it analytically} if one notices the relations
\begin{equation}\label{Eq22}
\arctan(1/\sqrt{3})={{\pi}\over{6}}\ \ \ \ \ \text{and}\ \ \ \ \ \arctan(\sqrt{3})={{\pi}\over{3}}\  .
\end{equation}
In particular, using Eq. (\ref{Eq22}) one finds from Eq. (\ref{Eq21}) 
that the critical density parameter, which marks the boundary between constant density 
stars that are characterized by the relation $T_{\text{c}}<T_{\text{s}}$ to compact 
stars that are characterized by the opposite relation $T_{\text{s}}<T_{\text{c}}$, is given 
by the dimensionless value
\begin{equation}\label{Eq23}
\Lambda^*=1\  .
\end{equation}

\section{Physical models in which it is always better to cross through the center of the compact star}

In the previous section we have explicitly proved that, for constant density stars, there is a 
critical density parameter [see Eqs. (\ref{Eq1}) and (\ref{Eq23})] above which the longer 
trajectory (the one that goes along 
the surface of the compact star) is characterized by a travel time which is shorter than the crossing 
time through the center of the star. 
In the present section we shall explicitly prove that there are physical models of compact stars in 
which the radial trajectory that passes through the center of the star is always (that is, for all allowed values 
of the physical parameters that characterize the star) characterized by the shorter travel time. 

To that end, we shall consider the physical model of thin-shell gravastars whose metric functions are 
characterized by the radial functional relations
\begin{equation}\label{Eq24}
A(r\leq R)=[B(r\leq R)]^{-1}=1-{{2Mr^2}\over{R^3}}\  .
\end{equation}
Substituting Eq. (\ref{Eq24}) into Eq. (\ref{Eq7}) one obtains the integral relation
\begin{equation}\label{Eq25}
T_{\text{c}}=\int^{R}_{0}{{2}\over{1-{{2Mr^2}\over{R^3}}}}dr\  .
\end{equation}
The integral (\ref{Eq25}) can be evaluated analytically to yield the dimensionless expression
\begin{equation}\label{Eq26}
{{T_{\text{c}}}\over{M}}=2^{{1\over2}}{\cal C}^{-{3\over2}}\tanh^{-1}\big(\sqrt{2{\cal C}}\big)\
\end{equation}
for the travel time through the center of the compact star. 

From the analytically derived expressions (\ref{Eq10}) and (\ref{Eq26}), which determine the compactness-dependent 
travel times between the two antipodal points of a star, one finds that, 
for the physical model of compact gravastars, the dimensionless 
ratio $T_{\text{c}}({\cal C})/T_{\text{s}}({\cal C})$ is a monotonically decreasing function 
of the dimensionless compactness parameter ${\cal C}$ of the star,
\begin{equation}\label{Eq27}
{{d\big({{T_{\text{c}}}\over{T_{\text{s}}}}\big)}\over{d{\cal C}}}<0\  ,
\end{equation}
with the asymptotic functional behaviors
\begin{equation}\label{Eq28}
{{T_{\text{c}}}\over{T_{\text{s}}}}\to {{2}\over{\pi}}<1\ \ \ \ \ \text{for}\ \ \ \ \ {\cal C}\to0\
\end{equation}
and \cite{Notelim2}
\begin{equation}\label{Eq29}
{{T_{\text{c}}}\over{T_{\text{s}}}}\to {{{2}\sqrt{1-2{\cal C}}}\over{\pi}}\to0\ \ \ \ \ 
\text{for}\ \ \ \ \ {\cal C}\to{1\over2}\  .
\end{equation}

Interestingly, from Eqs. (\ref{Eq27}), (\ref{Eq28}), and (\ref{Eq29}) one deduces that, for all physically allowed values of 
the dimensionless compactness parameter ${\cal C}\in(0,1/2)$ that characterizes the compact gravastars, 
the radial trajectory that connects the two antipodal points of the gravastar and 
passes through its center is characterized by the shorter travel time as measured by the remote operator (Alice).

\section{Summary}

The dimensionless density-area parameter $\Lambda\equiv 4\pi R^2\rho_{\text{max}}$ 
of compact stars is an important physical quantity that in principle can 
be bounded from below, using the general relativistic gravitational redshift effect, by far 
away observers \cite{Hodzz}. 
In the present compact paper we have explicitly demonstrated that the value of this density parameter may 
determine the preferred route to be taken by a spaceship whose remote operator (Alice) wants to send 
it from point A, which is located on the surface of a compact star, to its antipodal point B.  

Using the characteristic line element (\ref{Eq2}) of a compact star in a spherically symmetric curved spacetime, 
we have determined the travel times as measured by the 
remote operator for two different trajectories of the spaceship: 
a straight line that passes through the center of the star \cite{Notetun} 
and a semi-circular trajectory that goes along the surface of the star. 
Interestingly, we have explicitly proved that, similarly to what happens in cities with heavy traffic congestion, 
there is a critical density beyond which it is preferred to choose the longer path 
that connects the two antipodal points of the star along its perimeter. 

In particular, using the analytically solvable model of constant density stars, we have revealed 
the existence of the critical value [see Eqs. (\ref{Eq1}) and (\ref{Eq23})] 
\begin{equation}\label{Eq30}
\Lambda\equiv 4\pi R^2\rho=1\ 
\end{equation}
for the dimensionless density parameter of the star
above which the longer trajectory between the antipodal points (the one that goes along the surface of the star) is 
characterized by the {\it shorter} travel time as measured by the remote operator who is located 
far away from the star. 

It is worth noting that the analytically derived critical density parameter (\ref{Eq30}) corresponds 
to a constant density star whose dimensionless compactness parameter 
(mass-to-radius ratio) ${\cal C}\equiv M/R$ has the critical value
\begin{equation}\label{Eq31}
{\cal C}^*={1\over3}\  .
\end{equation}

Finally, we have explicitly proved, using the analytically solvable model of thin-shell gravastars, 
that there are physical situations in which the radial trajectory that passes 
through the center of the star 
is always (that is, for all physically allowed values of the parameters that characterize the compact stars) 
characterized by the shorter travel time between the two antipodal points of the compact star. 

%\newpage
\bigskip
\noindent {\bf ACKNOWLEDGMENTS}
\bigskip

This research is supported by the Carmel Science Foundation. I would
like to thank Yael Oren, Arbel M. Ongo, Ayelet B. Lata, and Alona B.
Tea for stimulating discussions.


\begin{thebibliography}{99}
\bibitem{Notetun} We consider an idealized scenario in which 
the remote controlled spaceship is capable of digging a thin tunnel through the star.
\bibitem{Noteunits} We use gravitational units in which $G=c=1$.
\bibitem{Notelmlm} Note that, for constant density stars, one can express the travel time (\ref{Eq10}) 
in terms of the dimensionless density-area parameter $\Lambda$ in the compact mathematical 
form ${{T_{\text{s}}}\over{M}}={{3\pi}\over{\Lambda\sqrt{1-{{2\Lambda}\over{3}}}}}$.
\bibitem{Notelim} Note that constant density stars are restricted 
to the dimensionless physical regime ${\cal C}\in(0,4/9)$, or equivalently 
$\Lambda\in(0,4/3)$ \cite{Buch}.
\bibitem{Notelim2} Note that thin-shell gravastars are restricted 
to the dimensionless physical regime ${\cal C}\in(0,1/2)$ [see Eq. (\ref{Eq24})].
\end{thebibliography}
\end{document}